\documentclass[aps,prd,nofootinbib]{revtex4}
\usepackage{amsmath,amssymb,pstricks,epsfig}
\usepackage{amsthm}
\usepackage[latin1]{inputenc}

\def\x{{\boldsymbol x}}

\def\p{{\boldsymbol p}}
\def\bs{\boldsymbol}

\theoremstyle{definition} 

\begin{document}

\title{\bf Pressure of massless hot scalar theory in the boundary effective theory framework}

\author{A. {\sc Bessa}$^{1}$,
\; F.T. Brandt$^{2}$, C.A.A. de Carvalho$^{3}$, E.S. Fraga$^{3}$}

\affiliation{$^{1}$Escola de Ci\^encias e Tecnologia, Universidade Federal do Rio Grande do Norte,
Caixa Postal 1524, 59072-970, Natal, RN, Brazil \\
$^{2}$Instituto de F\'\i sica, Universidade de S\~ao Paulo,
Caixa Postal 66318, 05315-970, S\~ao Paulo, SP , Brazil \\
$^{3}$Instituto de F\'\i sica, Universidade Federal do Rio de Janeiro,
Caixa Postal 68528, 21941-972, Rio de Janeiro, RJ , Brazil}

\date{\today}


\begin{abstract}
We use the boundary effective theory (BET) approach to thermal field theory in order to calculate the 
pressure of a system of massless scalar fields with quartic interaction. The method naturally separates 
the infrared physics, and is essentially non-perturbative. To lowest order, the main ingredient is the 
solution of the free Euler-Lagrange equation with non-trivial (time) boundary conditions.  We derive 
a resummed pressure, which is in good agreement with recent calculations found in the literature, following a 
very direct and compact procedure.
\end{abstract}

\maketitle

\section{Introduction}

In finite-temperature field theory it is known that a naive implementation of perturbation theory for 
the calculation of Feynman diagrams is ill-defined in the presence of massless bosons. This is due to 
the appearance of severe infrared divergences, brought about by the vanishing bosonic Matsubara mode 
in thermal propagators. These divergences plague the entire series, which becomes essentially 
meaningless \cite{FTFT-books}. 
As a result, one is forced to resort to resummation techniques that reorganize the perturbative series, 
and resum certain classes of diagrams, in order to extract sensible results. There are several ways of 
performing resummations and rewriting the degrees of freedom more efficiently in terms of quasiparticles; 
we refer the reader to the reviews \cite{Blaizot:2003tw,Kraemmer:2003gd,Andersen:2004fp} 
and to Ref. \cite{Bessa:2007vq} for a discussion and a list of specific references. 
All such techniques are designed to partially tame the infrared divergences, creating a non-zero domain 
of validity for weak-coupling expansions.
Nevertheless, the zero-mode problem remains, and the region of validity of resummed perturbative 
treatments can not be indefinitely enlarged.

In a recent paper \cite{bndeffaction}, we have proposed an alternative approach to thermal field 
theories, denoted by boundary effective theory (BET). The central idea of the method is to respect 
the double integral structure of the partition function in the functional integral formalism, 
\begin{subequations}\label{eq:Z}
\begin{align}
Z=\int[D\phi_0(\x)]\;\rho[\beta; \phi_0,\phi_0]\;,
\end{align}
where
\begin{equation}
\rho[\beta;\phi_0,\phi_0] = \int\limits_{\phi(0,\x)=\phi(\beta,\x)=\phi_0(\x)}[D\phi(\tau,\x)]\;
\;\;\;\; e^{-S[\phi]}\;
\end{equation}
\end{subequations}
is the diagonal element of the functional density matrix and $S[\phi]$ is the Euclidean action of the 
theory. This approach, based on the calculation of the density matrix, 
was already used in \cite{deCarvalho:2001xv} to construct dimensionally reduced effective actions, and
in \cite{Bessa:2007vq} to study the thermodynamics of scalar fields based on a semiclassical approximation.
The functional density matrix formulation of quantum statistics was discussed in \cite{Bessa:2008xj}.

As it will be clear in the sequence of this article, the double integration approach is essentially 
different from the one where a single functional integration over periodic configurations $\phi(\tau,\x)$ 
is performed. In the BET approach, the protagonist is the quantity $\phi_0(\x)$ \---- the field eigenvalue in the 
functional Schr\"odinger field-representation: $\hat{\phi}\,|\phi_0(\x)\rangle = \phi_0(\x)|\phi_0(\x)\rangle$. 
Indeed, any thermal observable can be constructed by integrating the appropriate functional of 
$\phi_0(\x)$ over the fields $\phi_0(\x)$ weighted by the corresponding diagonal element of the 
density matrix. The imaginary time evolution can be viewed as an intermediate step which calculates 
the weights for the effective theory of static $\phi_0$-fields. That effective theory defines a quantum 
statistical problem which encodes all information about thermalization; one is led to compute 
correlations of the field $\phi_0(\x)$, connecting the theory to physical quantities. 

The field $\phi_0(\x)$ has still another remarkable property: it is the zero (static) component of the 
dynamical field $\phi(\tau,\x)$. Indeed, to each dynamical configuration $\phi(\tau,\x)$  there 
corresponds a static configuration $\phi_0(\x) = \phi(0,\x) = \phi(\beta,\x)$.  We say that $\phi_0(\x)$ 
is the (time) boundary value of $\phi(\tau,\x)$. The difference $\phi(\tau,\x) - \phi_0(\x)$ vanishes at 
$\tau=0,\beta$, and so it can be expanded in a sine-Fourier series with non-zero frequencies 
$\hat{\omega}_n  = n\pi/\beta,\, n=1,2,\ldots$. From these considerations, we conclude that the 
effective theory encoded in $\rho[\beta;\phi_0,\phi_0]$ also contains all the infrared physics, and 
the double integral structure of $Z$ naturally separates the potentially divergent modes. 
 
The main result of Ref. \cite{bndeffaction} was the analytic calculation of the one-loop effective action 
for the boundary field in a scalar theory. Following the standard recipe, we used the saddle-point 
approximation, but in two steps: first, we kept $\phi_0(\x)$ fixed, and expanded the action around the 
classical configuration $\phi_c[\phi_0]$ \---- that has $\phi_0(\x)$ as its boundary value, and integrated 
over quadratic fluctuations vanishing at $\tau=0,\beta$; then, we expanded the resulting expression 
around the saddle-point $\phi_0 =0$, and took into account quadratic fluctuations of the boundary 
field. We showed that the one-loop effective action at finite temperature has the same expression as at 
zero temperature if written in terms of the classical field $\phi_c$ and if we trade free propagators at 
zero temperature by their finite temperature counterparts. Besides, we explained how to obtain a 
renormalized effective action in the case of a $\lambda \phi^4/4!$ theory. 

Now we address the problem of computing the pressure of the single-well quartic theory in the 
(problematic) massless limit and testing the aforementioned advantages of our framework. The 
weak-coupling calculation of the pressure for the massless hot scalar theory is an 
enterprise of about twenty years so far \cite{Parwani:1991gq,Frenkel:1992az,Parwani:1994zz,Braaten:1995cm,Gynther:2007bw}, with recent results to order $g^{8}\log g$ given in Ref. \cite{Andersen:2009ct}. 
Moreover, attempts to reorganize the perturbative series have followed different paths, generally 
introducing one or more variational parameters in the pressure, to be maximized in the end. Among 
these non-perturbative methods we find screened perturbation theory (SPT), introduced in thermal 
field theory in Ref. \cite{Karsch:1997gj}, which can also be implemented in the framework called 
optimized perturbation theory (OPT) presented in Ref. \cite{Chiku:1998kd}, the linear 
$\delta$-expansion (LDE) \cite{linear-delta}, and the so-called 2PI or $\Phi$-derivable methods \cite{2PI}.
All these methods, when applied to the thermodynamics of the scalar field, are remarkably more stable 
than the weak-coupling expansion (see Refs. \cite{Andersen:2000yj,Andersen:2008bz} for results 
using SPT, Refs. \cite{Pinto:1999py} for LDE, and 
Refs. \cite{Blaizot:1999ip,Peshier:2000hx,Braaten:2001en,Berges:2004hn} for 2PI). 

Due to the crucial role played by classical solutions $\phi_c[\phi_0]$ in the present approach (they 
are supposed to satisfy the Euler-Lagrange equation for arbitrary values of the coupling constant), 
the resulting pressure is essentially non-perturbative. Some technical difficulties arise, though. 
Decomposing the field as $\phi(\tau,\x) = \phi_c[\phi_0](\tau,\x) + \eta(\tau,\x)$, 
where $\eta(0,\x) = \eta(\beta,\x)=0$, 
introduces ultraviolet (UV) divergences in the calculation. In order to obtain a finite result, we use 
the renormalized effective action as derived in Ref. \cite{bndeffaction}. In addition, spurious UV 
divergences appear if one naively performs the saddle-point approximation and ignores higher order 
terms in the fluctuation $\eta(\tau,\x)$ and in the fluctuations of the boundary field. A finite expression 
for the pressure is obtained by taking into account the first term in the expansion of the self-interaction. 
The procedure to avoid UV divergences can, in principle, be extended, allowing for a systematic 
calculation of higher order corrections.

The structure of the paper is as follows: in Section \ref{fixedbnd}, we discuss the saddle-point 
approximation for the functional integration with fixed boundary configuration $\phi_0(\x)$, and how 
to implement corrections to that approximation; besides, we write the partition function in terms of the 
effective action for the boundary field; in Section \ref{bnffluctuation}, we perform the second functional 
integration using, again, the saddle-point method, and a renormalized expression for the pressure is 
obtained; finally, in Section \ref{conclusions}, we present our conclusions.

\section{Renormalized partition function in terms of the boundary field}\label{fixedbnd}

\begin{figure}[t!]
\begin{center}
\rotatebox{0}{%
     \resizebox{12cm}{!}{\includegraphics{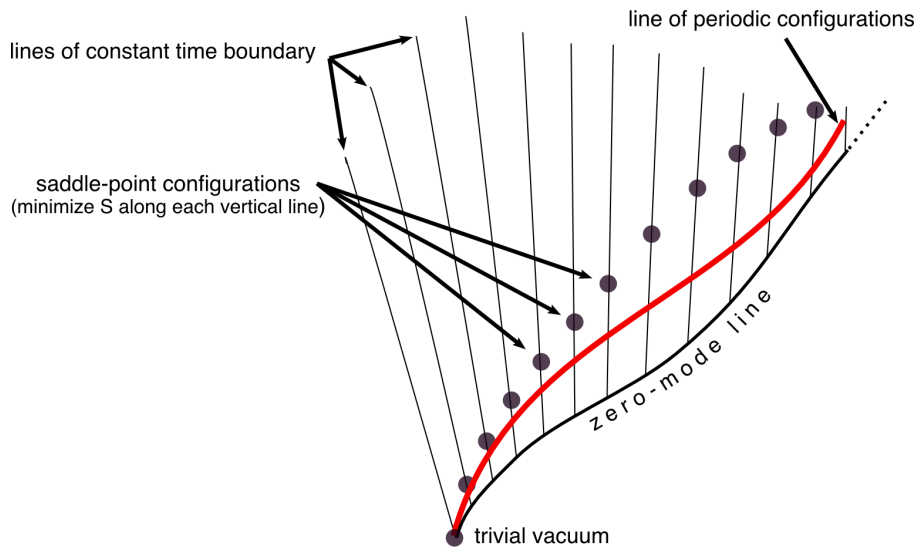}}}
\end{center}
\caption{\small Pictorial description of the partition of the functional domain of integration for $Z$ in the BET approach. The whole domain is indexed by the the zero-mode line. A given configuration $\phi(\tau,\x)$ lives on the vertical fiber over the boundary value $\phi(0,\x) = \phi(\beta,\x)$ in the bottom. Inside each vertical line, the Euclidean action is minimized by the saddle-point configuration (bullet). The line of periodic configurations is also shown.}\label{Fig:partition}
\end{figure}

Let us consider the Euclidean action,
\begin{equation}\label{euclideanaction1}
S[\phi]=
\int\limits_{0}^{\beta} (d^4x)_{_E}\, \left[\frac{1}{2}\partial_\mu \phi \partial^\mu\phi+ \frac{m_0^2}{2}\phi^2 +  U(\phi)\right]\;,
\end{equation}
where $(d^4x)_{_E}$ is a shorthand for $d\tau\,d^3x$. In this paper, $U(\phi) = \lambda \phi^4/4!$ and $m_0=0$. For single-well potentials like $U(\phi)$, the unique saddle-point of the action $S$ is the trivial vacuum $\phi =0$. However, in the density matrix approach, the functional domain of integration is partitioned in classes where all field configurations have the same time boundary. The restriction of $S$ to one of those classes (for instance, the one with boundary value $\phi_0(\x)$) has a non-trivial saddle-point $\phi_c[\phi_0]$, solution of
\begin{subequations}\label{eqmotion}
\begin{gather}
\square_{_E} \phi_c(x) + m_0^2 \phi_c(x) + U'(\phi_c(x)) = 0\;,\\\label{eqmotionb}
\phi_c(0,\x) = \phi_c(\beta,\x) = \phi_0(\x)\;,
\end{gather}
\end{subequations}
where $\square_{_E} = -(\partial_\tau^2 + \bs \nabla^2)$ is the Euclidean D'Alembertian operator and $x$ denotes $(\tau,\x)$. The first functional integration, being performed over configurations inside a certain class, will be dominated by fluctuations in the vicinity of the saddle-point of that class. In particular, the correspondence $\phi_0 \mapsto \phi_c[\phi_0]$ is 1:1 for single-well potentials. Therefore, one obtains an optimized spanning of the domain of integration in the calculation of $Z$ by collecting the contributions from quadratic fluctuations around a line of saddle-point configurations, as suggested by Fig. \ref{Fig:partition}. One is naturally led to a two-fold saddle-point approximation. It is worth remarking that $\phi_c(\tau,\x)$ is, in general, a non-periodic function of $\tau$ in the sense that $\partial_\tau\phi_c(0,\x) \neq \partial_\tau\phi_c(\beta,\x)$. One can verify it even in the simple case of a free theory. This is another important difference between BET and other methods.

The explicit dependence of $\phi_c$ on $\phi_0$ is not known, except in very special cases. In \cite{bndeffaction}, we obtained the following recursive relation for the classical solution,
\begin{eqnarray}
&&
\phi_c[\phi_0](\tau,\x)
=
\int d^3\x^\prime\;
\phi_0(\x^\prime)\;
\Big[
\partial_{\tau^\prime}G_0(\tau,\x;\tau^\prime,\x^\prime)
\Big]_{\tau^\prime=0}^{\tau^\prime=\beta}
\nonumber\\
&&
\qquad\qquad\qquad
-\int_{0}^{\beta}d\tau^\prime\int d^3\x^\prime\;
G_0(\tau,\x;\tau^\prime,\x^\prime)\;
U^\prime(\phi_c(\tau^\prime,\x^\prime))
\; ,
\label{eq:green1}
\end{eqnarray}
where $G_0$ is a Green function of the free operator,
\begin{subequations}\label{eq:bound_prop}
\begin{align}
\left (\square_{_E} +m_0^2\right )
G_0(x,x^\prime)
=\,\delta^{(4)}(x-x^\prime)\\
G_0(\tau,\x;0,\x^\prime)
=
G_0(\tau,\x;\beta,\x^\prime)
=0\; .
\end{align}
\end{subequations}
For a fixed boundary configuration, the fluctuations $\eta$ around $\phi_c$ vanish at $\tau =0$ and $\beta$,
\begin{gather}\nonumber
\phi(\tau,\x) = \phi_c(\tau,\x) + \eta(\tau,\x)\,\\\label{decomposephi}
\eta(0,\x) = \eta(\beta,\x) = 0\;.
\end{gather}
In terms of $\eta$, the renormalized partition function reads
\begin{align}
Z_R=\int[D\phi_0(\x)]
\int\limits_{\eta(0,\x)=\eta(\beta,\x)=0}[D\eta(\tau,\x)]
\;\;\;\; e^{-S[\phi_c +\eta] \;+\; \scriptsize\hbox{C.T}}\;,\label{eq:ZR}
\end{align}
where C.T. are counterterms to be chosen. One can think of each $\phi_c$ as a background field around which the dynamics of the fields $\eta$ takes place. 

In the vicinity of $\phi_c$, the action is approximately quadratic,
\begin{eqnarray}
S[\phi_c + \eta] \;=\; S[\phi_c] +\frac{1}{2}
\int (d^4x)_{_E}\;
\eta(x)\,\Big[\square_{_E} + m_0^2+ U^{\prime\prime}(\phi_c(x))\Big]\,\eta(x) \;+\;{\cal O}(\eta^3)\;.
\end{eqnarray}
It is convenient to introduce the Green function:
\begin{subequations}\label{eq:prop}
\begin{eqnarray}
\left[\square_{_E} + m_0^2 + U''\left(\phi_c(x^\prime)\right)\right]
G[\phi_c](x;x^\prime)
=\delta^{(4)}(x-x^\prime)\\
G[\phi_c](\tau,\x;0,\x^\prime)=G[\phi_c](\tau,\x;\beta,\x^\prime)=0\;. 
\end{eqnarray}
\end{subequations}
In particular, $G[0] = G_0$, the free propagator defined in Eq. \eqref{eq:bound_prop}. The Gaussian integration formally yields
\begin{equation}\label{saddleeta}
Z_R[\beta] \;\approx\; \int [{\cal D}\phi_0(\x)]\;e^{-S[\phi_c]\;+\; \scriptsize\hbox{C.T}} (\det G[\phi_c])^{1/2}\,.
\end{equation}
We should mention that the present saddle-point approximation can be good even in a strong-coupling regime. In fact, the classical solution is supposed to be exact for arbitrary values of the coupling constant. Besides, one can systematically improve the saddle-point approximation by expanding the cubic and quartic $\eta$-interactions, and calculating diagrams with lines of $G[\phi_c]$:
\begin{align}
Z_R[\beta] \;=\; \int [{\cal D}\phi_0(\x)]\,e^{-S[\phi_c]  \;+\; \scriptsize\hbox{C.T.}} (\det G[\phi_c])^{1/2} \,e^{-A[\phi_c]}\label{eq:ZR4}\;,
\end{align}
where
\begin{align}\label{eq:exp(-A)}
e^{-A[\phi_c]} = 1-\frac{\lambda}{8}\int (d^4x)_{_E} \,G^2[\phi_c](x,x) +  \frac{C_1}{2}\int (d^4x)_{_E}\,G[\phi_c](x,x) + {\cal O}(\hbox{3\,loops}) \;,
\end{align}
with $C_1$ being a mass counterterm.  The corresponding loop expansion is often called semiclassical series.

In \cite{bndeffaction}, we have shown that the renormalized 1-loop effective action for the boundary field $\phi_0$ is:
\begin{align}
\beta\,\Gamma_R[\phi_0] = S[\phi_c] + \frac{1}{2}\hbox{ Tr} \log\left( G^{-1}[\phi_c]\right)  + \frac{1}{2}\hbox{ Tr} \log\left( {\cal C}[\phi_c]\right)\;-\;\hbox{C.T.}\,,\label{gammaphi0a}
\end{align}
where
\begin{align}
{\cal C}[\phi_c] =  \frac{\delta^2 S[\phi_c]}{\delta \phi_0^2}\; = \;\left [\partial_\tau\partial_{\tau'}G[\phi_c]\right ]_0^\beta\label{fancyC}
\end{align}
is the 1-loop contribution from quadratic fluctuations of the field $\phi_0$.

 Comparing \eqref{eq:ZR4} and 
\begin{align}\label{expgammaphi0a}
e^{-\beta\,\Gamma_R[\phi_0]} = e^{-S[\phi_c] + \scriptsize\hbox{C.T.}}\; (\det G[\phi_c])^{1/2}\;(\det {\cal C}[\phi_c])^{-1/2}\;,
\end{align}
one can write
\begin{align}
Z_R[\beta] = (\det {\cal C}_0)^{1/2} \int [{\cal D}\phi_0(\x)]\,e^{-\beta\Gamma_R[\phi_0]} \;e^{-S_I[\phi_c]}\;,\label{eq:ZR6}
\end{align}
where ${\cal C}_0 = {\cal C}[0]$ and
\begin{align}
S_I[\phi_c] = A[\phi_c] -\frac{1}{2}\,\hbox{Tr}\log ( {\cal C}[\phi_c])  + \frac{1}{2}\,\hbox{Tr}\log ( {\cal C}_0)\;.
\end{align}
In this paper, the terms involving  $S_I$ and the interacting part of $S[\phi_c]$ will be dealt with order by order in a loop expansion that will be discussed in the next section.


\section{Pressure in the BET approach}\label{bnffluctuation}

Our strategy is to perform the second functional integration using again the saddle-point approximation. The saddle-point of $\Gamma_R[\phi_0]$ is the trivial configuration $\phi_0 =0$. Therefore, it is natural to expand $\Gamma_R$ in terms of the following $n$-point functions:
\begin{align}
\Gamma_R[\phi_0] = \sum_{n=1}^{\infty}\frac{1}{n!} \int \Gamma_R^{(n)}(\x_1,\ldots,\x_n)\,\phi_0(\x_1)\,\ldots\,\phi_0(\x_n)\, d^{3}\x_1\ldots d^{3}\x_n \;,
\end{align}
where
\begin{align}\label{gammaR}
\Gamma_R^{(n)}(\x_1,\ldots,\x_n) = \frac{\delta^{(n)}\Gamma_R[\phi_0]}{\delta \phi_0(\x_1) \ldots \delta \phi_0(\x_n)}\bigg |_{\phi_0=0}\;.
\end{align}

Up to quadratic order in $\phi_0$, we have (see Ref. \cite{bndeffaction}):
\begin{align}
\beta\,\Gamma_R[\phi_0] \approx \beta\,\Gamma_R^{(0)} + \frac{1}{2}\int \beta\,\Gamma_R^{(2)}(\x_1,\x_2)\,\phi_0(\x_1)\,\phi_0(\x_2)\, d^{3}x_1 \,d^3x_2\;,\label{gammaR_quad}
\end{align}
where $\Gamma_R^{(0)}= -V\,\pi^2/(90\beta^4)$ is essentially {\it minus} the pressure of an ideal gas of free massless bosons, V is the volume, and
\begin{align}
&\frac{\beta}{(2\pi)^3\delta(\p_1+\p_2)}\,\Gamma_R^{(2)}(\p_1,\p_2;\mu) =  2|\p_1|\,\tanh \frac{\beta |\p_1|}{2} + \beta\,m^2(|\p_1|;\beta)\;,\label{gammaR_saddle} 
\end{align}
with
\begin{align}\label{betam2}
\beta\,m^2(k;\beta) = \frac{\lambda}{24\beta}\frac{\tanh \beta k/2}{\beta k} \left (1 + \frac{\beta k}{\sinh \beta k} \right )\;.
\end{align}
The mass counterterm chosen was
\begin{align}
C_1 = \frac{\lambda}{2}\,\int^{\Lambda}\,\frac{d^4q}{(2\pi)^4}\,\Delta_F^0(q) \;,\label{massCT}
\end{align}
where $\Delta_F^0(q) = 1/q^2$ is the zero temperature (massless) free propagator in 4-dimensional Euclidean Fourier space. 

Substituting Eq. \eqref{gammaR_quad}  in \eqref{eq:ZR6} with $S_I=0$, and performing the quadratic integration over $\phi_0$, one obtains the saddle-point approximation for $Z$,
\begin{align}
Z_{sp}[\beta] = e^{-\beta\,\Gamma_R^{(0)}} (\det {\cal C}_0)^{1/2} (\det \beta \Gamma_R^{(2)})^{-1/2} = e^{\beta\,V P_{sp}}\;.\label{eq:Z_sp}
\end{align}
In Appendix A, we show that
\begin{align}\label{eq:P_sp}
P_{sp}(\beta) =  \frac{\pi^2 }{90\beta^4} - \frac{1}{2\beta} \lim_{\Lambda \rightarrow \infty}  \int^{\Lambda} \frac{d^3k}{(2\pi)^3} \, \log \left ( 1+ \beta m^2(k;\beta)\,\frac{\coth \beta k/2}{2k}\right) \;.
\end{align}

The second term on the r.h.s. of Eq. \eqref{eq:P_sp} can be identified with a series of daisy-diagrams, where the petal is given by \eqref{betam2}. As one can easily check, the ${\cal O}(\lambda)$ term in $P_{sp}$ is UV divergent. The 2-loop diagram carries the divergence. We will show that such a spurious divergence is removed when we consistently add the remaining 2-loop corrections to the saddle-point approximation. 

We have to be careful to identify the good propagator to represent the contraction of two $\phi_0$ fields. We know that such a propagator should be calculated at the saddle-point $\phi_0=0$. However, it is not obvious if the interacting mass should enter or not its definition. We claim that, in order to be consistent with the one-loop calculation of the effective action, we must use ${\cal C}_0$ defined in \eqref{fancyC} at $\phi_c=0$ as the propagator.

Notice that all quantities in this calculation depend on $\phi_0$ through $\phi_c$. Therefore, it is convenient to define the contraction of two fields $\phi_c$. In \cite{deCarvalho:2001xv}, it was shown that the propagator $\hat{\cal C}_0$ which is associated with that contraction satisfies
\begin{align}
\Delta_F\,=\,G_0\,+\,\hat{\cal C}_0\;,\label{calC0}
\end{align}
where $\Delta_F$ is the usual free thermal propagator, and $G_0$ is defined in Eq. \eqref{eq:bound_prop}. A pragmatic argument in favor of using lines of ${\cal C}_0$ (or $\hat{\cal C}_0$) to build 2-loop corrections is that it solves the problem of UV divergences of the saddle-point approximation, reproducing the correct result for the ${\cal O}(\lambda)$ perturbative contribution. We show that in Appendix B.

Finally, the renormalized pressure is given by 
\begin{align}\label{finalp}
P_{\scriptscriptstyle{BET}}(\beta)=  \frac{\pi^2 }{90\beta^4} -\frac{\lambda}{1152 \beta^4} - \frac{1}{2\beta} \lim_{\Lambda \rightarrow \infty}  \int^{\Lambda} \frac{d^3k}{(2\pi)^3} \,\bigg\{ \log \left ( 1+ \frac{\beta m^2(k;\beta)}{2}\frac{\coth \beta k/2}{k}\right) -  \frac{\beta m^2(k;\beta)}{2}\frac{\coth \beta k/2}{k} \bigg \} \;.
\end{align}

Figs. 2-4 plot the pressure normalized by the ideal pressure as a function of $g= \sqrt{\lambda}$ obtained using different methods. Fig. 2 compares the pressure from BET with that from weak-coupling calculations up to $g^8\,\log g$, according to Ref. \cite{Andersen:2009ct}. The weak-coupling expansion already includes resummation from order $g^3$ on. We see that the BET approach is in good agreement with the most recent results.

Figs. 3 and 4 compare BET with screening perturbation theory (SPT) calculations at two, three and four loops from Ref. \cite{Andersen:2008bz} over different ranges. In Fig. 3, the coupling $g$ goes from 0 to 8. It shows a complete mismatch between BET and the SPT curves for, say, $g \gtrsim 5$. However, in that range the SPT curves are not reliable either. In Fig. 4, we add the $g^8\,\log g$ weak-coupling curve to the comparison. Around $g = 3$, SPT curves present quite a large oscillation as one goes from  two to three, and then to four loops. We conclude that in the range where SPT shows convergence, the curve from BET seems to behave remarkably well.

\begin{figure}[t!]
\begin{center}
\rotatebox{0}{%
     \resizebox{12cm}{!}{\includegraphics{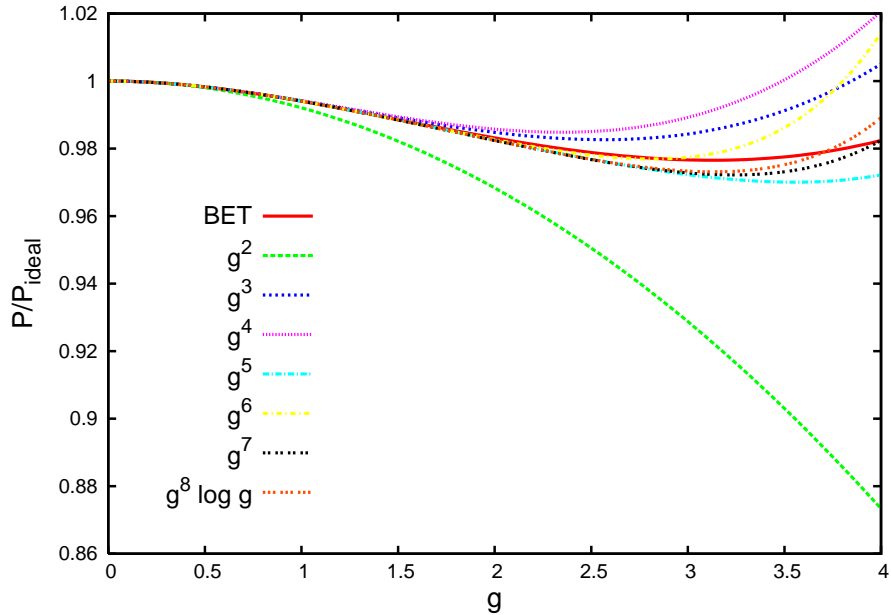}}}
\end{center}
\caption{\small Pressure normalized by $P_{ideal}$ as a function of the coupling constant $g=\sqrt{\lambda}$ in BET and weak-coupling (see \cite{Andersen:2009ct}) formalisms. The renormalization scale is $\mu = 2\pi T$. }
\end{figure}

\begin{figure}[t!]
\begin{center}
\rotatebox{0}{%
     \resizebox{12cm}{!}{\includegraphics{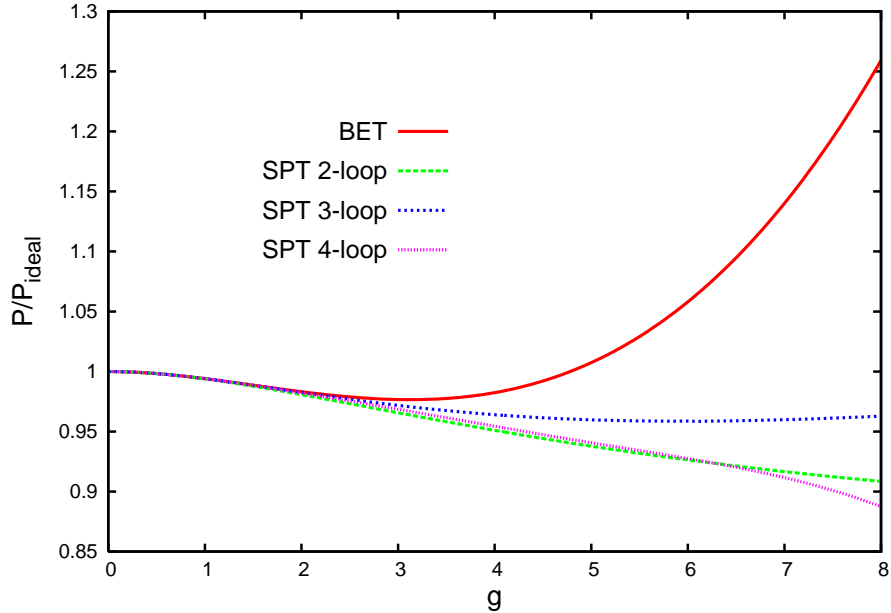}}}
\end{center}
\caption{\small Comparison of the normalized pressure obtained using BET and SPT calculations at two, three, and four loops reported in \cite{Andersen:2008bz}. The renormalization scale is $\mu = 2\pi T$. }
\end{figure}

\begin{figure}[t!]
\begin{center}
\rotatebox{0}{%
     \resizebox{12cm}{!}{\includegraphics{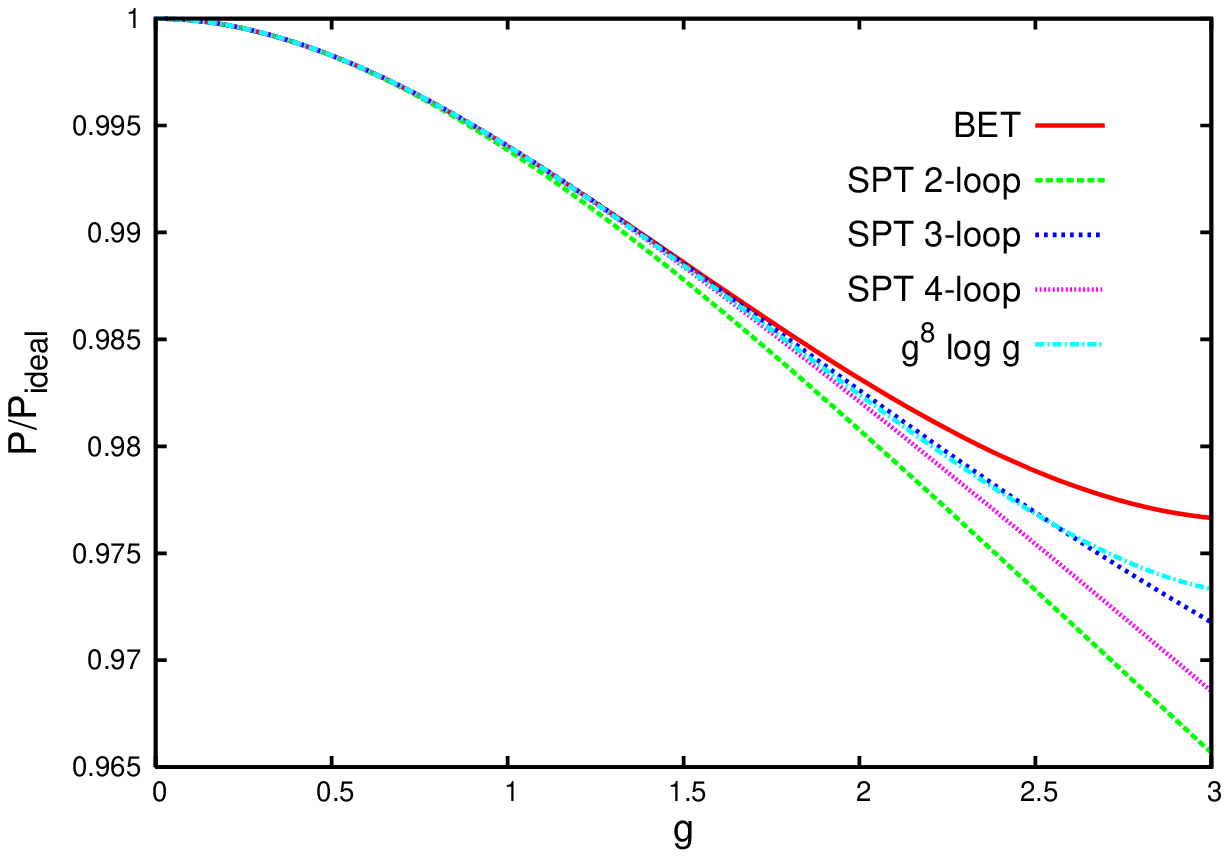}}}
\end{center}
\caption{\small Comparison of BET, $g^8\,\log g$ weak-coupling (Ref. \cite{Andersen:2009ct}) and SPT (Ref. \cite{Andersen:2008bz}) normalized pressures in the range $[0,3]$. The renormalization scale is $\mu = 2\pi T$. }
\end{figure}

\section{Conclusions}\label{conclusions}

The boundary effective theory (BET) was introduced as an alternative approach to quantum statistical mechanics. One of its main features is providing a natural separation of the zero mode (static) sector, leading to the construction of its effective theory which results from integrating over all the (imaginary time) dynamical modes. In previous work, the one-loop effective action for the zero mode had been calculated. In the present article, we have shown that BET is also a powerful method to attack a crucial problem of the thermodynamics of bosonic fields: infrared divergences.

The different strategies currently known to deal with such IR problems rely on some sort of resummation of diagrams of naive perturbation theory. The present calculation of the pressure using BET has the advantage of performing a highly nontrivial resummation in the scope of a natural and systematic procedure. In fact, the effective theory for the zero mode generated the whole series of daisy diagrams very naturally. Besides, in contrast with those built from thermal field theory, the daisy diagrams built in the context of BET fit quite well recent results in the literature obtained using rather involved techniques of screening perturbation theory at 4-loops. 

A distinctive feature of BET is the role played by field configurations which are not strictly periodic in the imaginary time. Indeed, following the double integral scheme for the partition function, we have shown that there is a line of saddle-point configurations dominating the functional integral and all of them, but the trivial one, are non-periodic.

We saw that the separation of the the field in its static and dynamical parts could lead to problems in the ultraviolet limit. This technical point has already been addressed in the calculation of the one-loop effective action. The solution is to perform a parallel calculation of the two functional integrations in the definition of the partition function. That led to a prescription for the present calculation which, we hope, can serve as a guide to extend the method to higher orders.

\acknowledgments

E.S.F. thanks J. O. Andersen and J. Pawlowski for fruitful discussions. The authors also 
thank J. O. Andersen and L. Kyllingstad for providing tables for their results for the pressure 
for comparison. 
This work was partially supported by CAPES, CNPq, FAPERJ, FAPESP and FUJB/UFRJ.


\appendix

\section{}

In \cite{bndeffaction}  we show that ${\cal C}_0(k) = 2k \tanh (k \beta /2)$. Expressing $\log \det {\cal C}_0$ as $\hbox{Tr} \log {\cal C}_0$, we obtain
\begin{align}
\log \,\det {\cal C}_0  = \int^{\Lambda} \frac{d^3k}{(2\pi)^3}\,\log\;\left [2k \tanh (k \beta /2)\right]\;.
\end{align}
Analogously, from \eqref{gammaR_saddle} it follows that
\begin{align}
\log\,\det \beta\,\Gamma_R^{(2)} =  \int^{\Lambda}\frac{d^3k}{(2\pi)^3}\,\log \;\bigg[ 2k\,\tanh \beta k/2 + \beta\,m^2(k;\beta) \bigg]\;.
\end{align}
Therefore,
\begin{align}
\log \left[\,(\det \,{\cal C}_0)^{1/2}\,(\det \beta\,\Gamma_R^{(2)})^{-1/2}\right] =  -\frac{1}{2}\int^{\Lambda} \frac{d^3k}{(2\pi)^3} \, \log \left ( 1+ \beta m^2(k;\beta)\,\frac{\coth \beta k/2}{2k}\right) \;.
\end{align}
Using that expression in \eqref{eq:Z_sp}, one obtains \eqref{eq:P_sp}.

\section{}

We start by expanding the interacting terms directly in \eqref{eq:ZR4}:
\begin{align}
e^{-\int (d^4x)_{_E} \,U(\phi)\, +\, C.T.} \approx 1-\frac{\lambda}{24}\,\int\,(d^4x)_{_E}\;\phi^4_c(x)\,+\,\frac{C_1}{2}\,\int \,(d^4x)_{_E}\,\phi_c^2(x)
\end{align}
and 
\begin{align}\label{eq:exp(-SI)at0}
e^{-A[\phi_c] } \approx 1-\frac{\lambda}{8}\int (d^4x)_{_E} \,G_0^2(x,x) + \frac{C_1}{2}\int (d^4x)_{_E}\,G_0(x,x)\;.
\end{align}
Contracting the fields $\phi_c$ using the proper symmetry factors, we obtain
\begin{align}
Z_{{\rm BET}}[\beta]\; \approx\; Z_{sp}[\beta]\;\bigg [ 1 -\frac{\lambda}{8}\,\int\,(d^4x)_{_E}\;\left (G_0^2(x,x) + \hat{\cal C}_0^2(x,x)\right )  + \frac{C_1}{2}\,\int \,(d^4x)_{_E}\,\left ( G_0(x,x) + \hat{\cal C}_0(x,x)\right )\bigg ]\;.
\end{align}
The 2-loop contribution from $Z_{sp}[\beta]$ is obtained expanding (see Ref. \cite{bndeffaction})
\begin{align}
(\det G[\phi_c])^{1/2} \approx (\det G_0)^{-1/2} \left [1- \frac{\lambda}{4}\,\int \,(d^4x)_{_E}\,G_0(x,x)\,\phi_c^2(x)\right ]\;,
\end{align}
and contracting the fields $\phi_c$. Collecting the 2-loop contributions for $P_{{\rm BET}} = (\log Z_{{\rm BET}})/\beta V$, we obtain
\begin{align}
P_{{\rm 2-loop}} =  -\frac{\lambda}{8\beta V}\,\int\,(d^4x)_{_E}\;\left (G_0(x,x) + \hat{\cal C}_0(x,x)\right )^2  + \frac{C_1}{2\beta V}\,\int \,(d^4x)_{_E}\,\left ( G_0(x,x) + \hat{\cal C}_0(x,x)\right )\;.
\end{align}
Using \eqref{calC0}, we have
\begin{align}
P_{{\rm 2-loop}} = -\,\frac{\lambda}{8\beta V}\,\int\,(d^4x)_{_E}\;\left (\Delta_F(x,x) -\frac{2C_1}{\lambda} \right )^2 + D\;,
\end{align}
where $D$ is a zero-temperature infinite constant which can be set to zero. Finally, using \eqref{massCT} and performing the remaining integration, we obtain that the ${\cal O}(\lambda)$ contribution to the pressure is finite and reproduces the perturbative result
\begin{align}
P_{{\rm 2-loop}} = -\frac{\lambda}{1152 \beta^4} \;.
\end{align}

\vspace{5mm}


\end{document}